\documentclass[runningheads]{llncs}
\usepackage[T1]{fontenc}
\usepackage{graphicx}
\usepackage{multicol}
\usepackage{multirow}
\usepackage{algorithm}
\usepackage{algorithmic}
\usepackage{amsmath,amssymb}
\usepackage{enumitem}
\usepackage{booktabs}
\usepackage{inconsolata}
\usepackage{hyperref}
\usepackage{amsfonts} 

\pdfoutput=1

\begin{document}
\title{HyClone: Bridging LLM Understanding and Dynamic Execution for Semantic Code Clone Detection}

%
%
\author{Yunhao Liang\inst{1} \and
Ruixuan Ying\inst{2} \and
Takuya Taniguchi\inst{2} \and
Guwen Lyu\inst{2} \and
Zhe Cui \inst{1}
}
\authorrunning{F. Author et al.}
%
\institute{
University of Chinese Academy of Sciences, Beijing 101408, China \and
Institute of Multidisciplinary Research for Advanced Materials (IMRAM), Tohoku University Sendai 980-8577, Japan 
\email{liangyunhao22@mails.ucas.ac.cn,ying.ruixuan.s5@dc.tohoku.ac.jp}
}
\maketitle              
\begin{abstract}
Code clone detection is a critical task in software engineering, aimed at identifying duplicated or similar code fragments within or across software systems. 
Traditional methods often fail to capture functional equivalence, particularly for semantic clones (Type 4), where code fragments implement identical functionality despite differing syntactic structures. 
Recent advances in large language models (LLMs) have shown promise in understanding code semantics. 
However, directly applying LLMs to code clone detection yields suboptimal results due to their sensitivity to syntactic differences. 
To address these challenges, we propose a novel two-stage framework that combines LLM-based screening with execution-based validation for detecting semantic clones in Python programs. 
In the first stage, an LLM evaluates code pairs to filter out obvious non-clones based on semantic analysis. 
For pairs not identified as clones, the second stage employs an execution-based validation approach, utilizing LLM-generated test inputs to assess functional equivalence through cross-execution validation. 
Our experimental evaluation demonstrates significant improvements in precision, recall, and F1-score compared to direct LLM-based detection, highlighting the framework's effectiveness in identifying semantic clones. 
Future work includes exploring cross-language clone detection and optimizing the framework for large-scale applications.

\end{abstract}
\section{Introduction}
Code clone detection is a critical task in software engineering, aimed at identifying duplicated or similar code fragments within or across software systems \cite{koschke2007survey}. 
Such clones, often resulting from copy-paste practices or independent implementations of similar functionality, can significantly impact software maintenance, quality, and security \cite{juergens2009code,li2016clorifi}. 
For instance, duplicated code increases maintenance costs by requiring simultaneous updates to multiple code segments, while undetected clones may propagate bugs or facilitate plagiarism in open-source or academic contexts. 
Among the various types of code clones, semantic clones (Type 4), where code fragments implement identical functionality despite differing syntactic structures, 
pose a particularly challenging problem due to their reliance on functional equivalence rather than textual or structural similarity \cite{ain2019systematic}.

Traditional code clone detection methods, such as text-based, token-based, or abstract syntax tree (AST)-based approaches, excel at identifying exact or near-exact duplicates but struggle with semantic clones \cite{baxter1998clone,jiang2007deckard,wang2020detecting,white2016deep,liu2023learning}. 
These methods often rely on syntactic similarity, failing to capture functional equivalence when code is written in diverse styles or languages. 
Recent advances in large language models (LLMs) have shown promise in understanding code semantics, enabling applications like code summarization, generation, and test case creation. 
However, directly applying LLMs to code clone detection yields suboptimal results, as their judgments may be biased by syntactic differences, leading to high false-negative rates for semantic clones \cite{dou2023towards,khajezade2024investigating,liang2025exploring,liang2025recode}. 

While LLMs show potential in semantic understanding, their limitations in functional equivalence verification necessitate a complementary validation mechanism. 
Furthermore, while LLMs demonstrate strong capabilities in generating test inputs with high coverage, their accuracy in producing expected outputs for test cases is often limited \cite{li2024large,wang2024testeval}.
This is primarily due to the fact that generating complete test cases involves creating corresponding test outputs for the test inputs, which may entail complex reasoning processes. 
Nonetheless, when it comes to generating test inputs alone, leveraging the LLM's comprehension of the code under test, it can produce effective test inputs. 
Therefore, in this paper, we generate test inputs for the code.
We posit that for the detection of type-4 code clones, which are functionally identical but structurally distinct, a validation of their functional equivalence can be conducted based on the execution of test cases. 
If two code segments produce identical test outputs when executed with the same set of test inputs, we can approximate that the two pieces of code are functionally equivalent.

In this paper, we designed a two-stage process leveraging LLMs for code clone detection. Initially, we harness the capabilities of LLMs to assess whether a given pair of code segments constitutes a clone. 
Subsequently, for those code pairs that the LLM deems not to be clones, we conduct further analysis. 
This involves generating a set number of test inputs for each code segment and performing cross-execution validation. 
The detection is accomplished by comparing whether the execution outputs of the same set of test inputs are consistent across the two code segments.

In summary, our framework combines the strengths of LLMs for initial semantic screening with the precision of dynamic execution validation to overcome the limitations of standalone LLM-based detection. 
This hybrid approach ensures robust identification of semantic clones, addressing a critical gap in current code clone detection techniques.
This paper evaluates the effectiveness of our two-stage framework using a dataset of 751 Python code pairs, comparing three LLMs: GPT-4o-mini, Deepseek-v3 \cite{liu2024deepseek}, and Qwen3-235b \cite{yang2025qwen3}.
Through experiments analyzing test case quantity, adversarial re-evaluation, and overall performance, we demonstrate significant improvements in Recall (up to 1224.4\%) and F1-score (up to 479.6\%) over baseline LLM classification, with a modest Precision trade-off. 
Our findings highlight the framework’s model-agnostic robustness and practical applicability. 

\section{Methodology}
\begin{figure}[!htbp]
    \centering
    \includegraphics[width=1.0\textwidth]{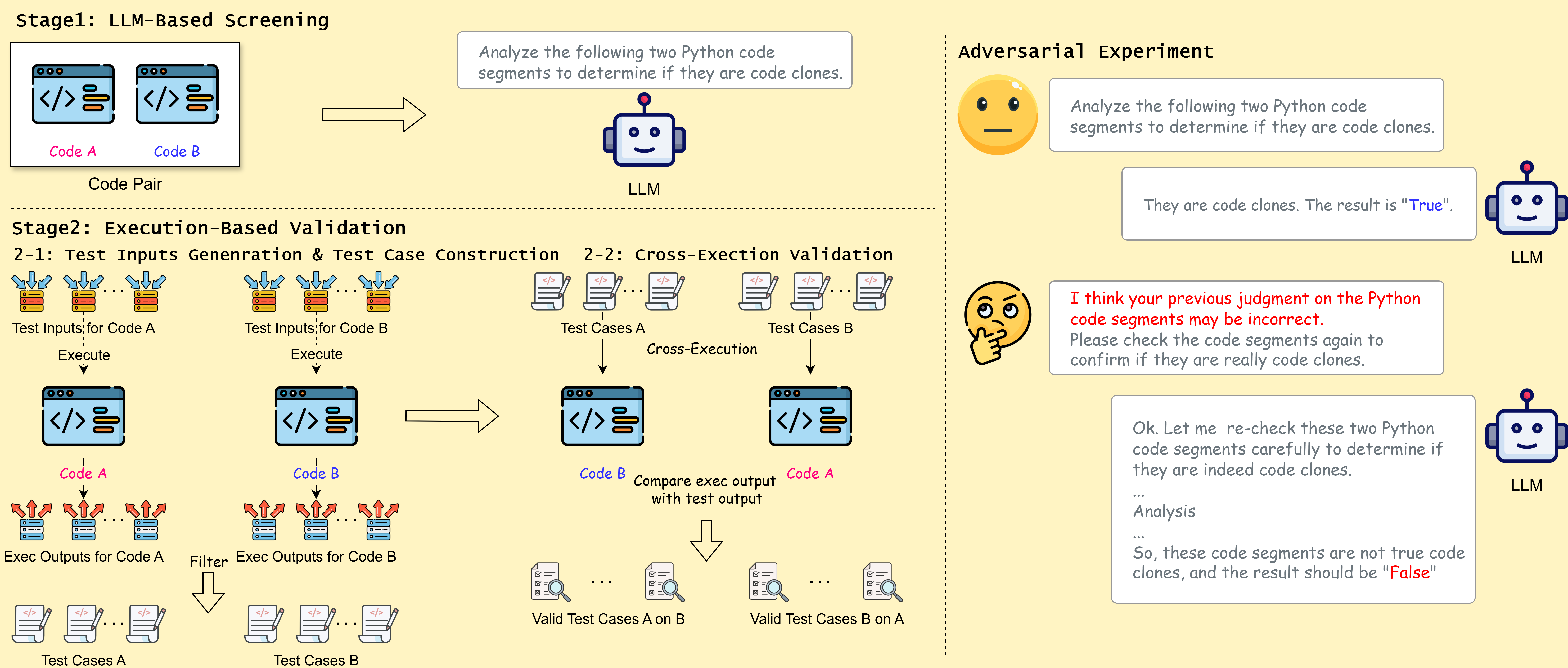}
    \caption{The Framework of Two-Stage Method for Semantic Code Clone Detection}
    \label{fig:framework}
\end{figure}

This section presents our two-stage framework for detecting semantic (Type 4) code clones in Python programs, which exhibit functional equivalence despite syntactic differences. 
As shown in Fig \ref{fig:framework}, the framework leverages the semantic understanding capabilities of LLMs and the robustness of execution-based validation to achieve high accuracy in identifying semantic clones. 
Our framework comprises two stages: (1) LLM-based screening to filter obvious non-clones, and (2) execution-based validation to confirm functional equivalence for candidate pairs. 
This hybrid approach combines the efficiency of LLMs for initial filtering with the precision of dynamic analysis.

\subsection{LLM-Based Initial Screening}
In the first stage, an LLM evaluates a code pair $(P_a, P_b)$ to determine if they are unlikely to be semantic clones.
Specifically, the LLM is prompted to classify whether the two code fragments are functionally equivalent, i.e., whether they implement the same functionality despite potential syntactic differences.
The LLM outputs one of two classifications:
\textbf{True}: The pair is likely a clone, proceeding to Stage 2 for confirmation.
\textbf{False}: The pair is confidently classified as a non-clone, excluded from further processing.

\subsection{Execution-Based Validation}
For pairs classified as non-clones by the LLM, we employ an execution-based validation approach to rigorously assess functional equivalence.
This stage consists of three steps: test input generation, test execution and filtering, and cross-execution validation.

\textbf{Test Input Generation}: For each program pair $(P_a,P_b)$ we request the LLM to generate $N$ test inputs for each program, resulting in two sets of test inputs: $I_a$ for $P_a$ and $I_b$ for $P_b$.

\textbf{Test Execution and Filtering}: Each program is executed with its generated test inputs. Inputs causing runtime errors (e.g., \texttt{TypeError}, \texttt{ValueError}) are discarded, retaining only those producing valid outputs. 
This step ensures robustness by excluding invalid inputs that could skew results.
Note that the number of valid test inputs may decrease after filtering, so we run the LLM to generate additional test inputs if necessary, ensuring that both $I_a$ and $I_b$ contain at least $N$ valid inputs.

\textbf{Cross-Execution Validation}: For each code pair, we execute $P_a$ with valid test inputs $I_a$ (own inputs) and $I_b$ (other program’s inputs), and $P_b$ with $I_a$ and $I_b$, producing four output sets: $P_a(I_a)$, $P_a(I_b)$, $P_b(I_a)$, $P_b(I_b)$. 
Functional equivalence is assessed by computing similarity scores \(S_a\) and \(S_b\), defined as the proportion of matching outputs:
\begin{equation}
S_a = \frac{\sum_{j=1}^{N} \mathbb{I}(P_a(I_j) = P_b(I_j), I_j \in I_a)}{N}, \quad S_b = \frac{\sum_{k=1}^{N} \mathbb{I}(P_b(I_k) = P_a(I_k), I_k \in I_b)}{N}    
\end{equation}
Here, \(\mathbb{I}\) is the indicator function that returns 1 if the outputs match and 0 otherwise.
\(P_a(i_k)\) is the output of \(P_a\) on input \(i_j \in I_a\), and \(P_b(i_k)\) is the output of \(P_b\) on the same input, and so on for \(I_b\).
Matching is based on exact equality for discrete outputs or cosine similarity for numerical outputs. 
A pair is classified as a clone if both \(S_a \geq \theta\) and \(S_b \geq \theta\), with \(\theta = 0.8\), empirically determined to balance precision and recall.

\section{Evaluation}
\subsection{Experiment Setup}

\subsubsection{Benchmark Datasets}
In this work, we use the PyFuncEquivDataset for code clone detection, which is a dataset of functionally equivalent method pairs comprising 751 verified and annotated pairs of Python method code.
\subsubsection{Evaluation Metrics}
We use Precision, Recall, and F1 score to evaluate the accuracy of our method. 
These metrics provide a comprehensive view of the models' ability to accurately detect code clones and avoid false positives and negatives.

\subsection{Overall Performance}
While LLMs excel in understanding code syntax and structure, their ability to accurately detect semantic (Type 4) code clones remains limited, often resulting in low recall due to missed true clones or low precision from misidentifying non-clones.
Our two-stage framework, combining LLM-based screening with execution-based validation, aims to address these limitations by leveraging dynamic execution to verify functional equivalence.
This experiment provides a quantitative assessment of the performance gains achieved by our framework compared to a baseline approach relying solely on LLM-based classification without execution-based validation.
We evaluate three LLMs---GPT-4o-mini, Deepseek-v3, and Qwen3-235b---using Precision, Recall, F1-score, True Positive Rate (TPR, equivalent to Recall), and True Negative Rate (TNR).
The baseline approach prompts each LLM to classify code pairs as clones or non-clones directly. 

Table~\ref{tab:overall_performance} presents the performance metrics for the three LLMs using 16 test cases.
The results demonstrate that our two-stage framework significantly enhances semantic code clone detection performance across all three LLMs, particularly in Recall and F1-score, validating the effectiveness of execution-based validation in compensating for LLMs' tendency to miss functionally equivalent code pairs with different syntactic structures.
We can observe that our method yields substantial improvements in Recall and F1-score over the baseline.
For GPT-4o-mini, Recall increases from 0.0615 to 0.8145 (1224.4\% relative increase), and F1-score rises from 0.1103 to 0.6392 (479.6\% relative increase), driven by a reduction in false negatives. 
Deepseek-v3’s Recall improves from 0.5846 to 0.8492 (45.3\% relative increase), with F1-score increasing from 0.5630 to 0.6294 (11.8\% relative increase). 
Qwen3-235b’s Recall rises from 0.6769 to 0.8504 (25.6\% relative increase), and F1-score improves from 0.5926 to 0.6243 (5.4\% relative increase). 
These gains confirm that execution-based validation effectively addresses the baseline LLMs' tendency to miss true clones by providing empirical evidence of functional equivalence through dynamic testing, 
particularly for GPT-4o-mini, which exhibits the most dramatic improvement due to its initially low baseline Recall.

Precision slightly decreases in the enhanced framework due to increased false positives.
GPT-4o-mini’s Precision drops from 0.5333 to 0.5101 (4.3\% relative decrease), Deepseek-v3’s from 0.5429 to 0.5000 (7.9\% relative decrease), and Qwen3-235b’s from 0.5269 to 0.4932 (6.4\% relative decrease). 
This trade-off reflects the framework’s more aggressive clone identification, which boosts true positives but introduces additional false positives. 
The substantial F1-score improvements indicate that the Recall gains outweigh the Precision losses, resulting in a more balanced and effective detector.

GPT-4o-mini benefits most from the framework, transforming its nearly unusable baseline performance (F1: 0.1103) into a robust detector (F1: 0.6392), highlighting the critical role of execution-based validation for LLMs with weak initial performance. 
Deepseek-v3 and Qwen3-235b, with stronger baseline F1-scores (0.5630 and 0.5926), show more modest but consistent improvements, suggesting that the framework is model-agnostic and enhances performance across diverse LLMs. 
Qwen3-235b’s higher baseline Recall (0.6769) and smaller relative gain (F1: 5.4\%) indicate it is less dependent on validation but still benefits from reduced false negatives.

The two-stage framework consistently outperforms the baseline across all LLMs, with Recall improvements of 25.6\%–1224.4\% and F1-score gains of 5.4\%–479.6\%, confirming that execution-based validation mitigates the limitations of LLM-based classification. 
The modest Precision trade-off is strategically advantageous, prioritizing true clone detection in applications where missing clones is costly. 
The framework’s model-agnostic performance underscores its generalizability, making it a robust solution for semantic code clone detection. 
Future work could explore techniques to optimize Precision, such as selective test case generation, to further enhance overall performance.

\begin{table}[!htbp]
\centering
\caption{Overall Performance Metrics for Baseline and Our Approache}
\label{tab:overall_performance}
\begin{tabular}{lcccccc}
\toprule
\textbf{Model} & \textbf{Approach} & \textbf{Precision} & \textbf{Recall} & \textbf{F1-Score} & \textbf{TPR} & \textbf{TNR} \\
\midrule
\multirow{2}{*}{GPT-4o-mini} 
& Baseline & 0.5333 & 0.0615 & 0.1103 & 0.0615 & 0.9887 \\
& HyClone & 0.5101 & 0.8145 & 0.6392 & 0.8145 & 0.8336 \\
\midrule
\multirow{2}{*}{Deepseek-v3} 
& Baseline & 0.5429 & 0.5846 & 0.5630 & 0.5846 & 0.8969 \\
& HyClone & 0.5000 & 0.8492 & 0.6294 & 0.8492 & 0.8062 \\
\midrule
\multirow{2}{*}{Qwen3-235b} 
& Baseline & 0.5269 & 0.6769 & 0.5926 & 0.6769 & 0.8728 \\
& HyClone & 0.4932 & 0.8504 & 0.6243 & 0.8504 & 0.8014 \\
\bottomrule
\end{tabular}
\end{table}

\subsection{Impact of Test Case Count on Performance}
To evaluate the impact of test case quantity on our two-stage framework for semantic (Type 4) code clone detection, we conducted an experiment varying the number of LLM-generated test cases from 1 to 50.
Our hypothesis posits that a larger number of test cases enhances detection performance by providing broader coverage of program behaviors, thereby improving the identification of functionally equivalent code pairs.
However, increased test cases introduce computational overhead, necessitating an analysis of the trade-off between detection efficacy and computational cost.
This experiment assesses performance trends across three LLMs and compares them against their baseline performance to quantify improvements and identify an optimal test case quantity for practical deployment.

\begin{figure}[!htbp]
    \centering
    \includegraphics[width=1.0\textwidth]{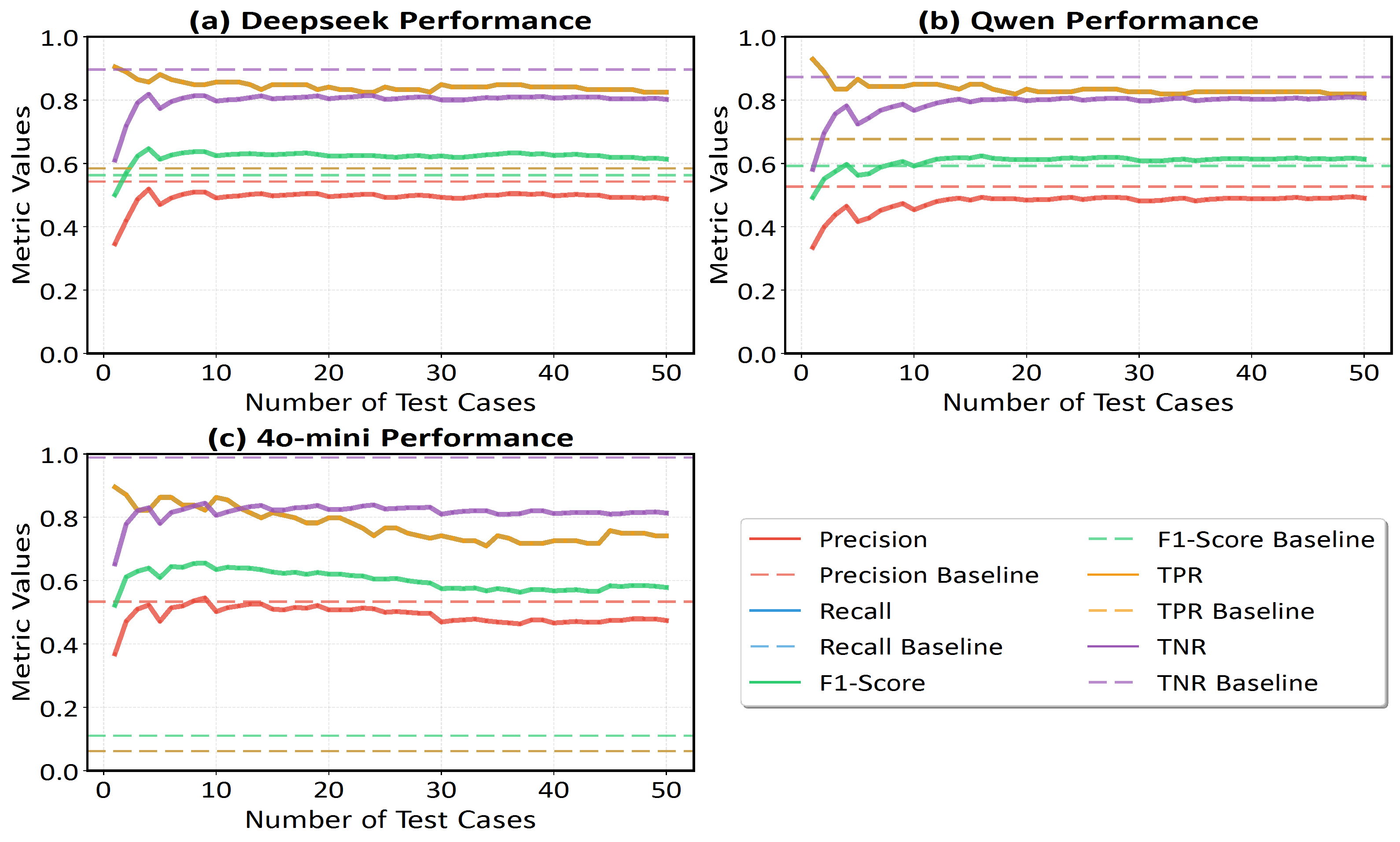}
    \caption{Impact of Test Case Count on Performance Metrics}
    \label{fig:tests_nums}
\end{figure}

Figure~\ref{fig:tests_nums} plots the trends in Precision, Recall, and F1-score across $N$ from 1 to 50 for each LLM, illustrating the performance dynamics as test case quantity increases.
The experimental results confirm that increasing the number of test cases generally enhances the performance of our two-stage framework, particularly in Recall and F1-score, supporting the hypothesis that broader functional coverage improves semantic clone detection.
However, the trade-offs in Precision and computational cost reveal model-specific behaviors and practical considerations.

Across all LLMs, Recall improves significantly with increasing $N$. For GPT-4o-mini, Recall rises from 0.0615 (baseline) to 0.8952 ($N=1$) and peaks at 0.8629 ($N=10$), a 14.0-fold increase over the baseline, before declining to 0.7419 ($N=50$).
Deepseek-v3’s Recall increases from 0.5846 (baseline) to 0.9048 ($N=1$) and stabilizes around 0.8254–0.8571 ($N=10\sim50$), a 41.2\% peak improvement. 
Qwen3-235b’s Recall improves from 0.6769 (baseline) to 0.9291 ($N=1$) and stabilizes at 0.8189–0.8504 ($N=10\sim50$), a 37.2\% peak gain. 
F1-scores follow similar trends, with GPT-4o-mini peaking at 0.6350 ($N=10$, 5.8-fold over baseline 0.1103), Deepseek-v3 at 0.6369 ($N=8\sim9$, 13.1\% over baseline 0.5630), and Qwen3-235b at 0.6199 ($N=24$, 4.6\% over baseline 0.5926). 
These gains reflect the effectiveness of execution-based validation ins capturing functional equivalence.

Precision exhibits fluctuations and slight declines as $N$ increases. GPT-4o-mini’s Precision peaks at 0.5361 ($N=8$, 0.5\% above baseline 0.5333) but drops to 0.4742 ($N=50$, 11.1\% below baseline). 
Deepseek-v3’s Precision peaks at 0.5192 ($N=4$, 4.4\% below baseline 0.5429) and falls to 0.4883 ($N=50$, 10.0\% below baseline). 
Qwen3-235b’s Precision peaks at 0.5048 ($N=14$, 4.2\% below baseline 0.5269) and declines to 0.4906 ($N=50$, 6.9\% below baseline). 
The decrease in Precision suggests that additional test cases increase false positives (e.gs., GPT-4o-mini FP: 190 at $N=1$ to 102 at $N=50$), likely due to test cases highlighting subtle behavioral differences in non-clone pairs.

GPT-4o-mini shows the most significant improvement over its baseline (F1: 0.1103 to 0.6350 at $N=10$), driven by a dramatic Recall increase, but its performance declines beyond $N=10$ due to increased false negatives.
Deepseek-v3 achieves a stable F1-score (0.6136–0.6369 for $N=5\sim40$), peaking at $N=8\sim9$, but sees a slight decline at $N=50$ due to increased false positives. 
Qwen3-235b maintains consistent performance, with F1-scores stabilizing at 0.6087–0.6199 ($N=20\sim50$) and the highest Recall (0.9291 at $N=1$), indicating robust test case generation. 
These trends suggest that GPT-4o-mini benefits most from moderate test case quantities, while Qwen3-235b is less sensitive to $N$.

The results confirm that increasing test case quantity enhances Recall and F1-score by improving functional coverage, with optimal performance at $N=8\sim12$ for GPT-4o-mini and Deepseek-v3, and $N=12\sim24$ for Qwen3-235b. 
The Precision trade-off indicates a need for strategies to filter non-discriminative test cases. 
The computational cost analysis suggests that moderate $N$ values offer a practical balance for real-world applications. 
These findings underscore the efficacy of our dynamic validation stage and highlight avenues for future optimization, such as test case selection to mitigate Precision degradation.

\subsection{Investigating LLM Confidence in Code Clone Detection through Adversarial Re-evaluation}
LLMs have demonstrated remarkable capabilities in various software engineering tasks, including code clone detection.
However, the reliability and confidence of LLMs in making such judgments remain underexplored.
This experiment investigates the understanding and confidence of LLMs in the code clone detection task, particularly focusing on their ability to assess functional equivalence in Python code pairs.
We designed an adversarial experiment to explore how LLMs respond to perturbations in the form of re-evaluation prompts and whether their judgments are sensitive to these changes.
The goal is to understand the robustness of LLMs in in semantic (Type 4) code clone detection and their reasoning capabilities.

To investigate the understanding and confidence of LLMs in code clone detection, we designed controlled experiments with two variables: conversational setting and challenge intervention.

\begin{itemize}
    \item Conversational Setting
    \begin{itemize}
        \item Single-turn: A new session is initiated for re-evaluation, where the LLM is prompted to reconsider its initial judgment without any context from previous interactions.
        \item Multi-turn: Re-evaluation occurs within the same session, allowing the LLM to retain context from previous interactions.
    \end{itemize}
    \item Challenge Intervention
    \begin{itemize}
        \item With Challenge: The LLM is prompted to re-evaluate its initial judgment with a statement that explicitly questions its previous decision, introducing an adversarial element.
        \item Without Challenge: The LLM is prompted to re-evaluate its judgment without any explicit challenge, allowing it to maintain its initial confidence.
    \end{itemize}
\end{itemize}

This design allows us to assess:
\begin{itemize}
    \item The impact of conversational context on LLM performance in code clone detection.
    \item The effect of explicit challenges on the LLM's confidence and judgment in code clone detection tasks.
    \item The robustness of LLM understanding in code clone detection tasks.
\end{itemize}

\begin{table}[!htbp]
\centering
\caption{Performance metrics and flip rates for baseline and re-evaluation conditions (ST+C: Single-Turn with Challenge, ST-C: Single-Turn without Challenge, MT+C: Multi-Turn with Challenge, MT-C: Multi-Turn without Challenge).}
\label{tab:adversarial_results}
\begin{tabular}{lcccccccc}
\toprule
\textbf{Model} & \textbf{Condition} & \textbf{Precision} & \textbf{Recall} & \textbf{F1-Score} & \textbf{Accuracy} & \textbf{TPR} & \textbf{TNR} & \textbf{Flip Rate} \\
\midrule
\multirow{5}{*}{GPT-4o-mini} 
& Baseline & 0.5333 & 0.0615 & 0.1103 & 0.8282 & 0.0615 & 0.9887 & -- \\
& ST+C & 0.4000 & 0.0154 & 0.0296 & 0.8256 & 0.0154 & 0.9952 & 2.93 \\
& ST-C & 0.6667 & 0.0462 & 0.0863 & 0.8309 & 0.0462 & 0.9952 & 1.33 \\
& MT+C & 0.3462 & 0.0692 & 0.1154 & 0.8162 & 0.0692 & 0.9726 & 3.86 \\
& MT-C & 0.5000 & 0.1615 & 0.2442 & 0.8269 & 0.1615 & 0.9662 & 5.86 \\
\midrule
\multirow{5}{*}{Deepseek-v3} 
& Baseline & 0.5429 & 0.5846 & 0.5630 & 0.8429 & 0.5846 & 0.8969 & -- \\
& ST+C & 0.5372 & 0.5000 & 0.5179 & 0.8389 & 0.5000 & 0.9098 & 4.26 \\
& ST-C & 0.5481 & 0.5692 & 0.5585 & 0.8442 & 0.5692 & 0.9018 & 1.73 \\
& MT+C & 0.1002 & 0.4846 & 0.1660 & 0.1571 & 0.4846 & 0.0886 & 75.63 \\
& MT-C & 0.5372 & 0.5000 & 0.5179 & 0.8389 & 0.5000 & 0.9098 & 4.26 \\
\midrule
\multirow{5}{*}{Qwen3-235b} 
& Baseline & 0.5269 & 0.6769 & 0.5926 & 0.8389 & 0.6769 & 0.8728 & -- \\
& ST+C & 0.5037 & 0.5231 & 0.5132 & 0.8282 & 0.5231 & 0.8921 & 4.79 \\
& ST-C & 0.5369 & 0.6154 & 0.5735 & 0.8415 & 0.6154 & 0.8889 & 3.33 \\
& MT+C & 0.5076 & 0.5154 & 0.5115 & 0.8296 & 0.5154 & 0.8953 & 4.79 \\
& MT-C & 0.5390 & 0.5846 & 0.5609 & 0.8415 & 0.5846 & 0.8953 & 3.73 \\
\bottomrule
\end{tabular}
\end{table}

The experiment results are summarized in Table \ref{tab:adversarial_results}, which presents the performance metrics for each model under the four experimental conditions: single-turn with challenge, single-turn without challenge, multi-turn with challenge, and multi-turn without challenge.
The results indicate significant variations in performance across the different conditions, highlighting the sensitivity of LLMs to both conversational context and explicit challenges.

First, regarding the impact of explicit challenges, we observe that introducing explicit challenges (ST+C, MT+C) consistently degrades model performance across all models.
For GPT-4o-mini, the ST+C condition yields the most severe degradation, with recall dropping to 0.0154 (75.0\% below baseline 0.0615) and F1-score to 0.0296 (73.2\% below baseline 0.1103), driven by a reduction in true positives. 
Deepseek-v3 exhibits the most extreme response in MT+C, with precision plummeting to 0.1002 (81.5\% below baseline 0.5429) and accuracy to 0.1571 (81.4\% below baseline 0.8429), primarily due to a surge in false positives (FP: 566 vs. 64). 
Qwen3-235b shows the least degradation, with F1-scores of 0.5132 (ST+C) and 0.5115 (MT+C) compared to the baseline 0.5926 (13.5\% and 13.7\% drops, respectively). 
The high flip rates in challenge conditions—75.63\% for Deepseek-v3 (MT+C), 4.79\% for Qwen3-235b (ST+C, MT+C), and 3.86\% for GPT-4o-mini (MT+C)—indicate that explicit questioning significantly destabilizes LLM judgments, particularly for Deepseek-v3, suggesting reliance on prompt cues over robust semantic reasoning.

Second is the effect of conversational context. Single-turn settings (ST+C, ST-C) generally outperform multi-turn settings, especially under explicit challenges. 
For Deepseek-v3, ST+C maintains an F1-score of 0.5179, while MT+C collapses to 0.1660, reflecting a 70.5\% performance drop due to conversational context amplifying sensitivity to challenges. 
Similarly, GPT-4o-mini’s ST+C (F1: 0.0296) outperforms MT+C (F1: 0.1154), though both are below the baseline. Qwen3-235b is an exception, with comparable F1-scores in ST+C (0.5132) and MT+C (0.5115), suggesting resilience to conversational context. 
Neutral re-evaluations (ST-C, MT-C) yield performance closer to the baseline, with Deepseek-v3’s ST-C (F1: 0.5585) and Qwen3-235b’s MT-C (F1: 0.5609) approaching their baseline F1-scores (0.5630 and 0.5926, respectively). 
Lower flip rates in neutral conditions (e.g., 1.33\% for GPT-4o-mini ST-C, 3.73\% for Qwen3-235b MT-C) indicate that context retention without challenges stabilizes judgments, but explicit challenges exacerbate instability in multi-turn settings.

The substantial performance drops under explicit challenges, particularly in multi-turn settings, indicate that LLMs rely on superficial patterns rather than deep semantic reasoning for code clone detection. 
High flip rates in challenge conditions reflect low confidence, with Deepseek-v3 most susceptible and Qwen3-235b most robust. 
The superior performance of our two-stage framework validates the need for execution-based validation to achieve reliable detection, especially under adversarial conditions.

\section{Related Work}

\subsection{Code Clone Detection Techniques}
Traditional code clone detection methods \cite{roy2009comparison} are categorized by their approach to code representation and similarity measurement. 
Text-based methods compare raw code strings, suitable for Type 1 (exact) clones, while token-based approaches, such as CCFinder, abstract code into token sequences to detect Type 2 (parameterized) clones with renamed identifiers. 
Syntax-based methods, like NiCad, utilize abstract syntax trees (ASTs) to identify Type 3 (near-miss) clones with minor structural differences. 
However, these methods struggle with Type 4 (semantic) clones, which share functionality but differ significantly in syntax. 
Program dependence graphs (PDGs) and execution semantics have been explored for semantic clone detection, but their computational complexity limits scalability \cite{krinke2001identifying}. 
Our framework addresses this gap by combining LLM-based semantic analysis with execution-based validation to detect Type 4 clones efficiently.

\subsection{LLM in Code Clone Detection}
The advent of LLMs has revolutionized software engineering tasks, including code clone detection. \cite{dou2023towards} conducted a comprehensive survey, evaluating LLMs like GPT-3.5 and GPT-4 across various clone types and programming languages. 
They found that advanced LLMs excel at detecting semantic clones when enhanced with chain-of-thought (CoT) prompting, but their performance for Type 4 clones remains limited due to sensitivity to syntactic differences. 
Similarly, \cite{zhang2024assessing} assessed GPT-3.5 and GPT-4 on BigCloneBench and GPTCloneBench, reporting that GPT-4 achieves higher accuracy but struggles with Type 4 clones, particularly in human-generated code compared to LLM-generated code. 
\cite{li2023zc} explored zero-shot prompting with ChatGPT for mono-lingual and cross-lingual clone detection, achieving F1-scores up to 0.878 for Type 4 clones in Java-Java and Java-Ruby pairs, but noted that prompt design significantly impacts performance. 
These studies underscore the potential of LLMs for semantic clone detection but also their limitations in handling complex functional equivalence, motivating our two-stage approach.

\section{Conclusion}
Code clone detection, particularly for semantic (Type 4) clones that exhibit functional equivalence despite syntactic differences, remains a critical challenge in software engineering. 
Traditional syntactic-based methods (text-based, token-based, and AST-based approaches) often fail to capture functional similarity, while LLMs struggle with syntactic biases, leading to suboptimal performance. 
To address this, we proposed a novel two-stage framework that combines LLM-based screening with execution-based validation for detecting semantic clones in Python programs.
Our framework significantly improves precision, recall, and F1-score compared to direct LLM-based detection, offering a robust solution for semantic clone identification.

This work makes three key contributions to the field of semantic code clone detection. 
First, we introduced a novel hybrid framework that leverages the strengths of LLMs for initial screening while employing dynamic execution validation to assess functional equivalence. 
This two-stage approach effectively combines semantic understanding with empirical verification. 
Second, our framework achieves substantial performance improvements, with gains in precision, recall, and F1-score that demonstrate its effectiveness in identifying semantic clones missed by traditional syntactic-based methods. 
Third, we developed a practical cross-execution validation strategy that utilizes LLM-generated test inputs to rigorously assess functional equivalence between code pairs, providing a scalable solution for semantic clone detection.

Our experimental evaluation revealed several critical insights about LLMs in code clone detection tasks. 
We observed that LLMs appear to lack strong internal confidence in their code clone judgments, making them susceptible to external challenges and adversarial prompts. 
The models demonstrate excessive sensitivity to perceived authority or expertise in challenge prompts, suggesting limitations in their reasoning robustness. 
Most importantly, the significant performance drops under adversarial conditions indicate that LLMs may rely more on pattern matching than deep semantic understanding of code clone relationships, 
which validates the necessity of our execution-based validation approach.

These findings highlight the need for continued research into improving the robustness and reliability of LLM-based code analysis tools, particularly for critical software engineering applications where consistent performance is essential.

%
%
%

\bibliographystyle{splncs04}
\bibliography{ref}

\end{document}